\documentclass[11pt, a4paper, conference, twocolumn]{IEEEtran}
\IEEEoverridecommandlockouts

\usepackage{tikz}
\usepackage{subcaption}

\usepackage{placeins}
\usepackage{cite}
\usepackage{amsmath,amssymb,amsfonts}
\usepackage{algorithmic}
\usepackage{graphicx}
\usepackage{textcomp}
\usepackage{doi}
\usepackage{float}
\usepackage{xcolor}
\usepackage{orcidlink}
\usepackage[margin=1.27cm]{geometry} 
\graphicspath{{./images/}}

\definecolor{ieeeblue}{RGB}{0,51,102} 
\usepackage{hyperref}
\hypersetup{
    colorlinks=true,
    linkcolor=ieeeblue,     
    urlcolor=ieeeblue,
    citecolor=ieeeblue
}
\usepackage[T1]{fontenc}
\usepackage[scaled=0.92]{helvet}

\usepackage[fontsize=11pt]{scrextend}

\def\BibTeX{{\rm B\kern-.05em{\sc i\kern-.025em b}\kern-.08em
    T\kern-.1667em\lower.7ex\hbox{E}\kern-.125emX}}

\begin{document}

\title{\textbf{TERA}: A Unified \textbf{T}aylor Model \textbf{E}nabled \textbf{R}eachability \textbf{A}nalysis Framework}

\author{
    \IEEEauthorblockN{Salma Iraky\IEEEauthorrefmark{1}\thanks{$^*$Corresponding author (\texttt{s.iraky@lancaster.ac.uk}). Both authors are with the School of Computing and Communications, Lancaster University, UK. This work was presented at \textsf{\href{https://web.archive.org/web/20260622121207/http://control2026.ukcontrol.org/wp-content/uploads/2026/06/Control-2026-Proceedings-Booklet.pdf}{CONTROL~2026}}: 15th United Kingdom Automatic Control Council (UKACC) International Conference on Control, 23–25 June 2026. 
    } and Andrew Sogokon}
    }

\maketitle

\begin{abstract}
    Reachability analysis of safety-critical systems requires computing rigorous enclosures of all possible state trajectories. 
    Taylor Model (TM)-based methods have proved effective at mitigating the so-called \emph{wrapping effect} which leads to overly conservative enclosures of reachable sets. However, existing tools are often hard to extend or focused on narrow system classes (e.g. deterministic systems modelled by ODEs, or hybrid systems). 
    We develop \textsf{TERA}: a \textsf{Python}-native framework for TM-based reachability analysis of continuous, hybrid and stochastic systems within a single symbolic-numeric workflow.
    \textsf{TERA} is free and open-source, enabling rapid prototyping of reachability analysis techniques with rigorous enclosures.
    At present, our implementation is able to compute tight reachable set over-approximations for non-linear ODEs and hybrid systems on difficult benchmark problems, and already supports analysis of continuous-time stochastic systems. Our goal is to develop a robust open-source \textsf{Python} infrastructure for rigorous reachability analysis supporting a broad class of systems, including stochastic hybrid systems.
\end{abstract}

\section{Introduction and Motivation}
In a system, e.g. given by $\dot{x} = f(x,u,w)$, where $u$ denotes a control input and $w$ represents bounded disturbances or uncertainty, a feedback law $u=\kappa(x)$ induces the closed-loop dynamics $\dot{x} = f(x,\kappa(x),w)$.
If the system is safety-critical, it is important to establish whether, e.g., all closed-loop trajectories remain within prescribed bounds (and perhaps verify other safety requirements).
Given a set of initial states $X_0$ and a finite time horizon $T>0$, the \emph{forward reachability problem} is to compute sets $\{\mathcal{R}(t)\}_{t \in [0,T]}$ that contain all closed-loop trajectories originating in $X_0$ (with probabilistic guarantees when under stochastic disturbances).
In practice, computing these so-called \emph{flowpipes} exactly is only possible for some special classes of \emph{linear} systems, so one is forced to resort to set-based \emph{over-approximations} of system trajectories over a finite time horizon; if these provide a rigorous enclosure of the reachable set, one can perform (bounded time) safety verification of the closed loop system.

Common set representations for over-approximating flowpipes include (hyper-)intervals and zonotopes; however, reachability analysis based on these representations suffers from the~\emph{wrapping effect}, where accumulated over-approximations rapidly yield very conservative enclosures of the reachable set. 
Taylor Models (TMs; see~\cite{berz1998verified}) offer an alternative set representation that mitigates this phenomenon by preserving functional dependencies through high-order polynomial representations with rigorous remainder bounds, making them especially well-suited for reachability analysis of nonlinear systems. 
Numerous tools have been developed for reachability analysis using TMs, such as \textsf{Flow$^*$} (in \textsf{C++}), \textsf{CORA} (in \textsf{MATLAB}) and \textsf{JuliaReach} (in \textsf{Julia}). However, these are either less suited to rapid prototyping, require a proprietary environment to run, or do not (currently) handle stochastic systems.

\textsf{TERA} is a fully \textsf{Python}-native free and open-source reachability framework designed to conveniently integrate with the scientific \textsf{Python} ecosystem and draws on the functionality offered by the \textsf{SageMath} computer algebra system.\footnote{\textsf{TERA} is available from \href{https://github.com/sssabry/tera}{\texttt{github.com/sssabry/tera}} and is distributed under the terms of the \textsf{GPLv2} license.} 
Currently, \textsf{TERA} is capable of computing rigorous TM-based enclosures of reachable sets for continuous systems described by \emph{non-linear ODEs} (this includes systems with non-polynomial nonlinearities, including transcendental functions such as $\sin$ and $\cos$, appearing in the right-hand side), \emph{hybrid dynamical systems} that combine discrete and continuous behaviour and in which the dynamics of continuous modes may be governed by non-linear ODEs, and \emph{stochastic systems} (currently support only extends to continuous-time systems).
\section{Methodology}
Within \textsf{TERA}, reachable sets are represented using TMs (see~\cite{berz1998verified,hybridnonlinear_chen}) of the form $P(x_0,t)+I$, where $P$ is a Taylor polynomial approximation (up to some degree)  of the solution $x(x_0,t)$ to a system of ODEs and $I$ is an interval such that the true solution is guaranteed to be enclosed within the TM.\footnote{i.e. $x(x_0,t) \in I + P(x_0,t)$. We should note that computing such an enclosure requires \emph{correct rounding}, which is achieved using the \textsf{GNU MPFR} library (integrated within \textsf{SageMath}).}
Continuous-time dynamics are propagated using validated TM integration schemes. Specifically, \textsf{TERA} supports local single-step integration grounded in the foundational work of Berz and Makino on validated ODE solving (e.g. see~\cite{berz1998verified} and \cite{hybridnonlinear_chen}), together with the flowpipe construction approach for non-linear continuous systems presented by Chen in~\cite{hybridnonlinear_chen}. 
To mitigate the wrapping effect over long time horizons, \textsf{TERA} additionally supports compositional left-right propagation following the so-called \emph{shrink-wrapping} method introduced by Berz and Makino and later refined by B\"unger \cite{DBLP:journals/na/Bunger18}.

\begin{figure*}[ht]%
\centering
\begin{subfigure}{.65\columnwidth}
  \centering
  \includegraphics[width=\linewidth]{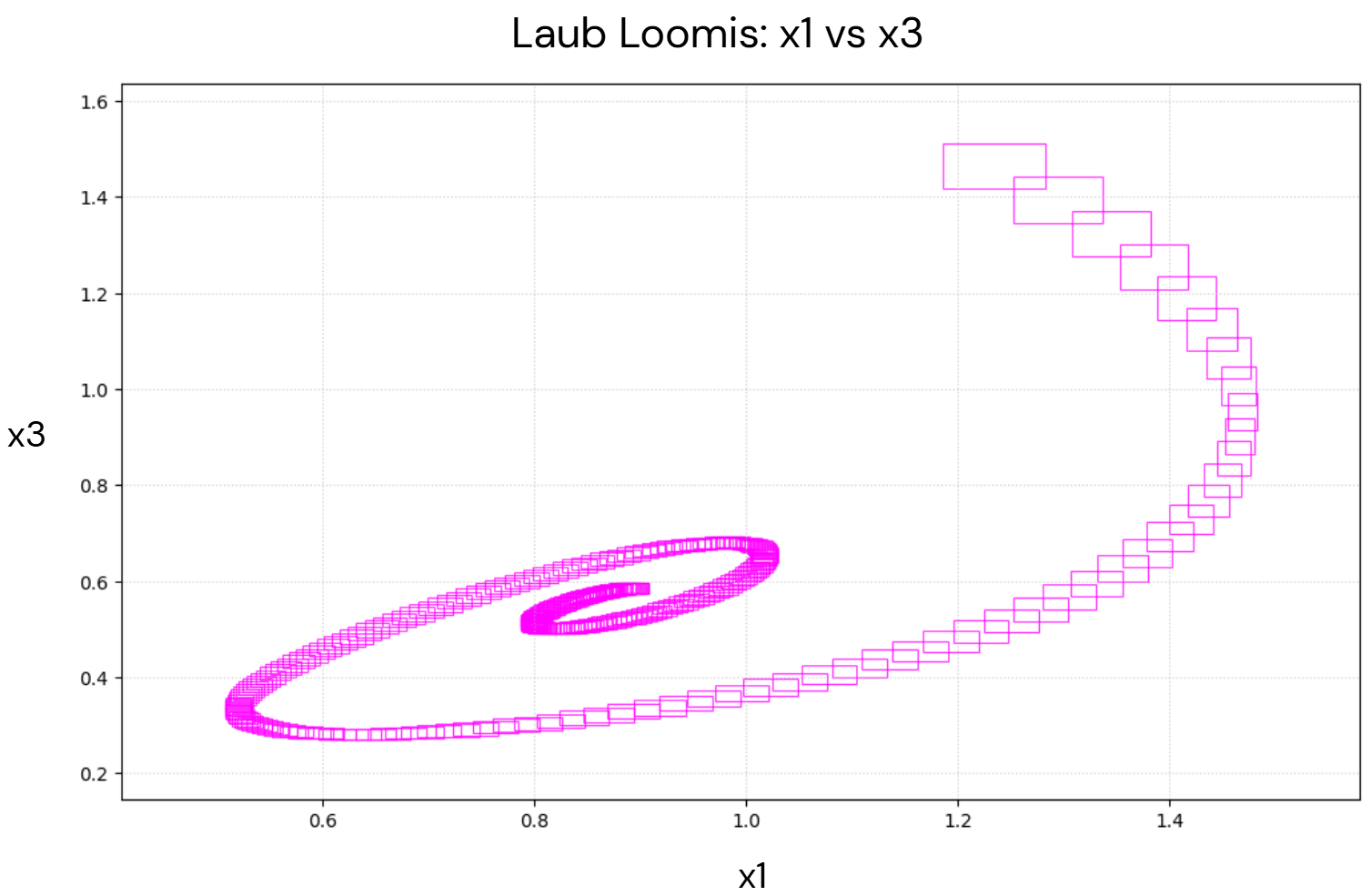}
  \caption{Non-linear Continuous System}
  \label{subfig:nonlinear}
\end{subfigure}\hfill
\begin{subfigure}{.65\columnwidth}
  \centering
  \includegraphics[width=\linewidth]{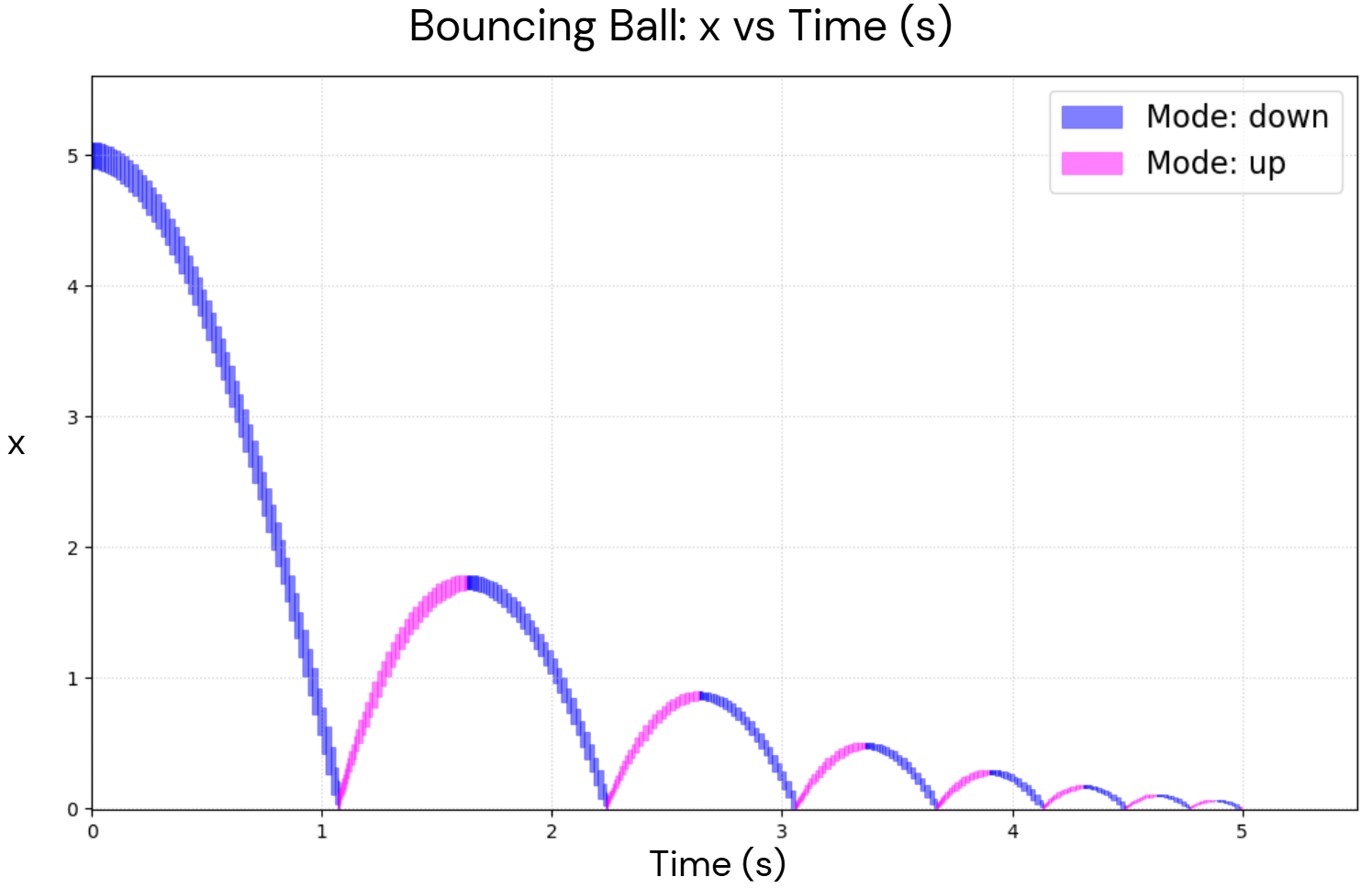}
  \caption{Hybrid System}
  \label{subfig:hybrid}
\end{subfigure}\hfill
\begin{subfigure}{.65\columnwidth}
  \centering
  \includegraphics[width=\linewidth]{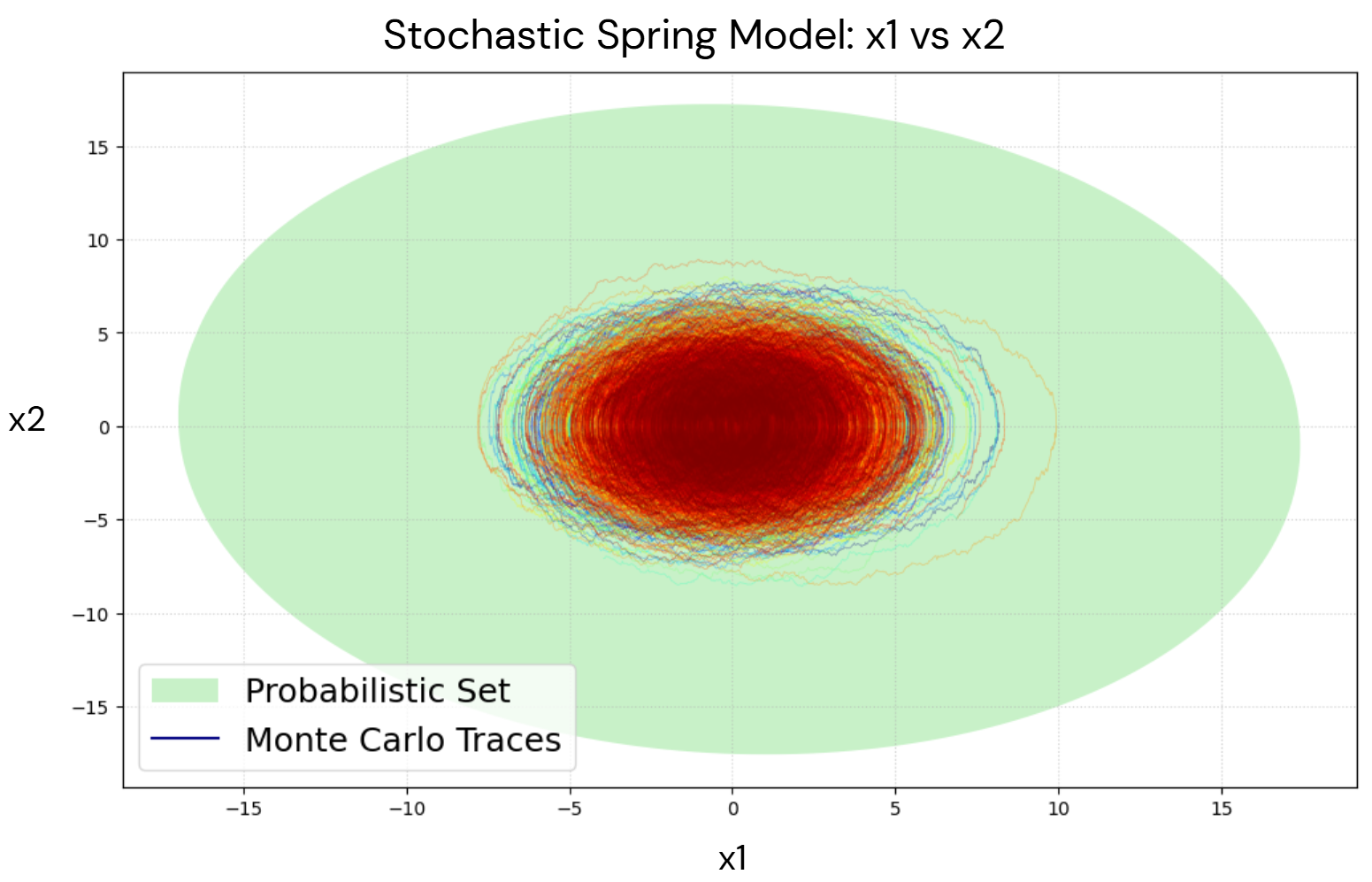}
  \caption{Stochastic System}
  \label{subfig:stochastic}
\end{subfigure}

\caption{Reachable set enclosures produced by \textsf{TERA}} 
\label{fig:plots}
\end{figure*}

\textsf{TERA} supports hybrid dynamics, computing hybrid automaton guard intersections and discrete transitions following the TM-based hybrid reachability semantics described by Chen in~\cite{hybridnonlinear_chen}. 
Current support for continuous-time stochastic dynamics is achieved by combining deterministic TM flowpipes with probabilistic deviation bounds. Following the recent work of Jafarpour et al.~\cite{stochasticreach_jafarpour}, \textsf{TERA} computes $\delta$-probabilistic reachable set ($\delta$-PRS) enclosures, guaranteeing that system trajectories remain within the computed sets with probability at least $1-\delta$. 

\section{Illustrative Examples}
We validated \textsf{TERA}'s current capabilities against the \textsf{ARCH} competition benchmark problem suite \cite{arch2018}, featuring continuous and hybrid  system verification problems. 
Fig.~\ref{subfig:nonlinear} shows a plot of the reachable set enclosures produced by \textsf{TERA} (projected onto the $(x_1,x_3)$ plane) for the $7$-dimensional non-linear Laub-Loomis biochemical reaction network, whose dynamics is given by: 
\begin{small}
\[
\begin{cases}
\dot x_1 = 1.4 x_3 - 0.9x_1, \quad
\dot x_2 = 2.5 x_5 - 1.5 x_2, \\
\dot x_3 = 0.6 x_7 - 0.8 x_2x_3, \quad
\dot x_4 = 2 - 1.3x_3x_4, \\
\dot x_5 = 0.7 x_1 - x_4x_5, \quad
\dot x_6 = 0.3 x_1 - 3.1 x_6, \\
\dot x_7 = 1.8 x_6 - 1.5 x_2x_7.
\end{cases}
\]
\end{small}
\textsf{TERA} computes a tight enclosure of the reachable set over the time horizon $t \in [0,20]$ from a small hyper-rectangular initial set in 13 seconds of computation time on this problem.\footnote{Using a machine with a 3.3 GHz 8-core \textsf{Snapdragon X Plus} CPU and 16 GB of RAM. Each rectangle in the figure represents a bound on the TM enclosure of the reachable set $\{\mathcal{R}(t)\}$ from the initial set over an integration time-step interval \mbox{$t\in[t_i,t_{i+1}]$}.}
\textsf{TERA} also scales to larger non-linear benchmark problems with transcendental functions in~\cite{arch2018}, such as the $12$-dimensional controlled quadrotor problem.

Fig.~\ref{subfig:hybrid} illustrates the rigorous enclosures computed by \textsf{TERA} for the bouncing ball hybrid system~\cite[Ex. 4.1.2]{hybridnonlinear_chen} from initial displacement $x(0) \in [4.9,5.1]$ and initial velocity $v(0) \in [-0.2,0]$ over the time horizon $t \in [0,5]$. Figure ~\ref{subfig:hybrid} shows the projection of the resulting flowpipes on the $(x, t)$ plane, coloured to reflect the active mode of the hybrid automaton (\textsf{up}, \textsf{down}). 

Fig.~\ref{subfig:stochastic} illustrates an example featuring a stochastic spring model from~\cite[Ex. 6.6]{applied_SDEs} with parameters $\nu=1$, $\gamma=\frac{1}{10}$ and $\beta(t)$ being standard scalar Brownian motion: 
\begin{small}
\[d\mathbf{x} =
\begin{pmatrix} 0 & 1 \\ -\nu^2 & -\gamma \end{pmatrix}\mathbf{x}\,dt
\;+\;
\begin{pmatrix} 0 \\ 1 \end{pmatrix}\,d\beta(t),
\qquad \mathbf{x}=\begin{pmatrix}x_1\\x_2\end{pmatrix}\,.
\]
\end{small}
 \textsf{TERA} computes an enclosure of the $\delta$-PRS over the horizon $t \in [0,20]$ where $\delta = 0.001$, shown in green in Fig.~\ref{subfig:stochastic} where it is seen to contain $2000$ Euler-Maruyama Monte Carlo sample paths from the initial set.

\section{Conclusion and Future Work}



\textsf{TERA} is the first \textsf{Python}-native free and open-source framework for Taylor Model-based reachability analysis of continuous, hybrid and stochastic systems. Our current implementation already supports non-linear ODEs, hybrid systems and continuous-time stochastic systems and is able to solve difficult benchmark problems.\footnote{ The \textsf{ARCH} competition benchmark problems and other examples of running \textsf{TERA} are available on  {\scriptsize \texttt{\href{https://github.com/sssabry/tera/tree/main/Examples}{\textsf{GitHub}}}} along with documentation and usage instructions.}

Ongoing work is focused on further extending \textsf{TERA}'s capabilities to support \emph{stochastic hybrid systems}\cite{DBLP:journals/automatica/LavaeiSAZ22}, with the long-term goal of developing a robust open-source \textsf{Python} infrastructure for rigorous reachability analysis that can be applied yet more broadly, e.g., to systems controlled by artificial neural networks.


\bibliographystyle{IEEEtran}
\bibliography{references}

\end{document}